\documentclass[journal]{IEEEtran}

\ifCLASSINFOpdf

\else

\fi

\usepackage{cite}
\usepackage{amsmath,amssymb,amsfonts}
\usepackage{graphicx}
\usepackage{booktabs}
\usepackage{array}
\usepackage{url}
\usepackage{hyperref}
\usepackage{times}
\usepackage[utf8]{inputenc}
\usepackage[compatibility=false]{caption}  
\usepackage{float}
\usepackage{lipsum}
\usepackage{amsmath,amsthm,amssymb,amsfonts}
\usepackage{makecell}
\usepackage{comment}
\usepackage{multirow}
\usepackage{rotating, graphicx}
\usepackage{xcolor}
\usepackage{tabularx}
\usepackage{orcidlink}
\usepackage{ragged2e}
\usepackage{textcomp}
\usepackage{listings}
\usepackage{algorithm}
\usepackage{algpseudocode}
\usepackage{setspace}
\usepackage[normalem]{ulem}
\useunder{\uline}{\ul}{}
\usepackage{hyperref}
\usepackage{xcolor}
\usepackage{booktabs} 
\usepackage[utf8]{inputenc}
\usepackage{cite}

\begin{document}

\title{Machine Learning Techniques for Enhancing Quantum Key Distribution}

\author{
    Ali Al-Kuwari\IEEEauthorrefmark{1}\textsuperscript{\orcidlink{0009-0007-2312-5921}} (alal55457@hbku.edu.qa),
    Safaa Alqrinawi\IEEEauthorrefmark{1}(saal88803@hbku.edu.qa),
   Lujayn Al-Amir\IEEEauthorrefmark{1} (lual88833@hbku.edu.qa),
 Amina Mollazehi\IEEEauthorrefmark{1} (ammo68489@hbku.edu.qa),

    and 
    Saif Al-Kuwari\IEEEauthorrefmark{1}\textsuperscript{\orcidlink{0000-0002-4402-7710}} (smalkuwari@hbku.edu.qa),

\IEEEauthorblockA{\IEEEauthorrefmark{1}Qatar Center for Quantum Computing, College of Science and Engineering, Hamad Bin Khalifa University, Doha, Qatar}\\
}

\maketitle

\begin{abstract}
Quantum Key Distribution (QKD) offers theoretically unbreakable security by leveraging the principles of quantum mechanics. However, its practical implementation is challenged by environmental vulnerabilities, noise and hardware imperfections. Recently, Machine Learning (ML) has emerged as a powerful tool to address these limitations and enhance the real-world viability of QKD systems. In this survey, we provide a thorough review of the ML techniques applied to improve QKD security and performance. These ML techniques are typically applied to five applications. First, parameter optimization,  focusing on key tasks such as signal calibration, polarization alignment, phase stabilization, modulation state tuning, and post-processing enhancements, all aimed at maximizing secure key generation and minimizing error rates. Second, attack detection where ML models are used to identify and classify quantum threats, such as photon-number-splitting and Trojan-horse attacks. Third, protocol selection, which leverages ML to dynamically choose or switch QKD protocols based on operational conditions, enhancing adaptability and resilience. Fourth, key performance prediction including predictive modeling of core performance metrics such as Secret Key Rate (SKR) and Quantum Bit Error Rate (QBER). Finally, quantum network management, where ML is applied at the system level to optimize large-scale QKD deployments through intelligent routing, node management, and resource allocation. This structured framework provides a comprehensive view of how ML supports both physical-layer optimization and high-level system intelligence in QKD. 
Performance improvements are evaluated using metrics such as accuracy, reduced QBER, and increased SKR. While ML has demonstrated significant potential in securing QKD systems for critical applications in finance, government, and defense, challenges remain in scalability, computational demands, and real-world testing. 
Ongoing work should consider developing lightweight, generalizable models and establishing standardized evaluation benchmarks to enable practical deployment of ML-enhanced QKD in real-world quantum cryptographic infrastructures.
\end{abstract}


\section{Introduction}
Quantum physics introduces unique properties, such as entanglement, superposition, the uncertainty principle, the no-cloning theorem, and the concept of quantum measurement. These properties are the foundation of quantum cryptography, which is a promising solution to rising threats against classical cryptographic systems. Traditionally, cryptographic systems have relied on mathematical problems that are hard to solve with classical computers. However, the advent of quantum computing raises serious concerns for these systems as they can be broken by quantum computers using algorithms like Shor's algorithm \cite{Shor1994}. In response, Post-Quantum Cryptography (PQC) has been proposed, offering quantum-resistant algorithms based on new mathematical approaches, such as lattice-based, code-based, hash-based, or multivariate polynomial schemes \cite{kumar2021state}. However, PQC still depends on mathematical assumptions that may not be provably secure.

The quantum solution to this threat is Quantum Key Distribution (QKD), which offers a different approach by relying on the laws of quantum mechanics rather than mathematical complexity. QKD uses quantum properties, such as the no-cloning theorem, which makes it impossible for an attacker to copy quantum states without being detected \cite{imran2024quantum}. This gives QKD a significant advantage over classical key exchange methods, as any eavesdropping attempt will inevitably introduce detectable errors in the quantum channel. Consequently, today, QKD is widely considered suitable for high-security applications such as finance, defense, and critical infrastructure \cite{gyongyosi2019survey}.

Although QKD is theoretically secure, practical implementations still face many real-world challenges caused by hardware limitations and environmental noise. These attacks can exploit hardware vulnerabilities in the system without always being detected. To address these challenges, recent research has applied Machine Learning (ML) techniques to enhance QKD performance and security. ML techniques can help automate complex decision-making tasks, making them well-suited for adaptive control in dynamic environments. For example, ML has been applied to tasks such as anomaly detection, noise filtering, system parameter optimization, and secure key rate (SKR) prediction \cite{imran2024quantum}. Supervised learning models, such as Support Vector Machines (SVMs) and neural networks (NNs), have been used to detect known attacks based on the behavior of the system \cite{tunc2023machine,mao2020detecting, al2021machine}. Deep learning (DL) approaches help in phase stabilization and polarization correction, especially in dynamic environments \cite{zhou2024adjusting, ahmadian2022cost}. Reinforcement learning methods have been proposed to adapt the parameters of the QKD protocol to maximize key rates under varying conditions, such as satellite channels \cite{ahsun2025adaptive}. These efforts show that ML can help reduce the quantum bit error rate (QBER), increase the SKR, and improve robustness across most of the QKD protocols.

Recent literature demonstrates a growing interest in leveraging ML to enhance the practicality and security of QKD as shown in the comparison Table~\ref{tab:qkd_ml_comparison_litrature}. 
Huang \emph{et al.} \cite{huang2021secure} investigated the integration of ML within continuous-variable QKD (CV-QKD) systems, where ML models such as SVMs, random forests (RFs), and neural networks are employed for real-time attack detection, system parameter prediction, and modulation optimization. This approach not only improves anomaly detection but also reduces the dependency on physical monitoring hardware, making CV-QKD systems more adaptable to real-world imperfections. Mafu \cite{mafu2024advances} provided a broader review of AI and ML in quantum communication, emphasizing the role of supervised learning models in optimizing parameters and enhancing SKRs, particularly in MDI-QKD. The study highlights ML’s value in replacing computationally expensive simulations and enabling adaptive device calibration under noisy conditions. Similarly, Long \emph{et al.} \cite{long2023survey} offered a comprehensive survey focused on ML-assisted CV-QKD, showcasing how deep learning and regression techniques contribute to excess noise filtering, reconciliation accuracy, and real-time key rate estimation, thereby reinforcing the protocol's robustness in dynamic environments. In contrast, Chandre \emph{et al.}\cite{article} took a broader view across quantum cryptography as they focused on exploring ML across various quantum cryptographic domains, including QKD, quantum secure direct communication (QSDC), and quantum networks, highlighting applications such as error correction, adaptive protocol selection through reinforcement learning (RL), and ML-based cryptanalysis for proactive security.

Compared to the studies summarized in Table ~\ref{tab:qkd_ml_comparison_litrature}, our review offers a more comprehensive and structured taxonomy by classifying ML applications in QKD into five key categories: parameter optimization,  quantum attack detection, protocol selection, key performance prediction, and quantum network communication. This refined framework enables us to capture both low-level improvements such as signal calibration and system tuning and high-level functionalities, including adaptive protocol control and network-wide optimization. While previous reviews often focus on specific protocols (e.g., CV-QKD) or narrow technical challenges (e.g., key rate maximization), our work provides a broader and more integrated perspective. It maps ML techniques across a wide range of QKD use cases and highlights their contributions within a unified structure.

\begin{table*}
  \caption{Comparison of ML applications in QKD across selected surveys.}
  \label{tab:qkd_ml_comparison_litrature}
  \centering
  \small
  \renewcommand{\arraystretch}{1.2}
  \begin{tabular}{@{}l l p{0.2\linewidth} p{0.2\linewidth} l@{}}
    \toprule
    \textbf{Paper} & \textbf{Scope} & \textbf{ML use cases} & \textbf{Security impact} & \textbf{Techniques} \\
    \midrule
    \cite{mafu2024advances} & General quantum communication & Parameter optimization & Device/channel robustness & RF, NN \\
    \cite{long2023survey}   & CV-QKD focused               & Noise filtering \& reconciliation & System-wide adaptability & DL, regression, classification \\
    \cite{article}          & Broad (QKD, QSDC, quantum networks) & Error correction \& key optimization & Protocol-level defense & RL, NN, clustering \\
    \cite{huang2021secure}  & CV-QKD focused               & Attack detection \& parameter tuning & Targeted attack mitigation & SVM, RF, NN, kNN \\
    \bottomrule
  \end{tabular}
\end{table*}

This paper presents a comprehensive review of ML applications in QKD systems, organized into five main domains: parameter optimization, quantum network management, attack detection, protocol selection, and key performance prediction. This classification offers a structured perspective on how ML contributes to enhancing both the physical-layer performance and system-level intelligence of QKD. By analyzing the existing literature through this framework, we identify the strengths, limitations, and research gaps when applying ML to real-world QKD implementations. Unlike previous surveys that focus on specific protocols or isolated tasks, our review covers multiple QKD protocols, including discrete-variable (DV-QKD), continuous-variable (CV-QKD), Measurement-Device-Independent (MDI-QKD), and Twin-Field (TF-QKD), and spans the full QKD workflow, from individual components to network-scale deployments, providing details and discussions of ML’s role in advancing secure and performance of quantum communication systems.

The remainder of this paper is organized as follows. The Background section introduces the fundamentals of QKD, with an overview of four protocols: DV-QKD, CV-QKD, MDI-QKD, and TF-QKD. The Taxonomy section defines the categorization logic adopted in this review. The ML for QKD Parameter Optimization section is divided into protocol-specific and protocol-agnostic strategies, including optimization of polarization, phase stability, modulation variance, and post-processing procedures. It also includes application-specific optimization efforts. The next section focuses on ML for QKD Network Optimization, followed by ML for QKD Attack Detection, ML for QKD Protocol Selection, and ML for QKD Key Metrics Prediction. The paper concludes with a summary of current challenges, research gaps, and future directions.

\section{Preliminaries}
\label{sec:background}

QKD enables two authorized parties (Alice and Bob) to establish secret keys with security grounded in quantum effects. Any eavesdropping attempt disturbs the system and is detectable, consistent with the No-Cloning Theorem \cite{gyongyosi2019survey}. With the growing risk that large-scale quantum computers pose to asymmetric schemes, QKD is increasingly viewed as a candidate for high-assurance environments in finance, government, healthcare, and defense \cite{liu2019practical}. At a high level, protocols are grouped into DV-QKD \cite{Djordjevic2019} and CV-QKD \cite{cvDjordjevic2019}.

\subsection{Discrete Variable QKD (DV-QKD)}
States are prepared at the single-photon level and measured with single-photon detectors, yielding strong security guarantees against interception \cite{Djordjevic2019}. This family includes both prepare-and-measure and entanglement-based designs.
\begin{itemize}
    \item {\texttt{Prepare-and-Measure}}: Individual photons are emitted in specific bases and analyzed by the receiver. In BB84, two bases (rectilinear/diagonal) with four polarization states are used; a subset comparison exposes tampering. B92 uses two non-orthogonal states, reducing outcomes while still supporting key generation under suitable conditions \cite{win2023analysis}.
    \item {\texttt{Entanglement-Based Protocols}}: Keys are extracted from correlated measurement outcomes on shared entangled pairs, with Bell tests validating the correlations. In BBM92, a source distributes one photon of each pair to Alice and the other to Bob; matching bases are kept, mismatches are discarded, and security follows from the observed correlations \cite{win2023analysis}.
\end{itemize}

\subsection{Continuous Variable QKD (CV-QKD)}
Information is encoded in field quadratures using coherent or squeezed states and read out via homodyne/heterodyne detection, which fits well with standard telecom hardware. Widely studied variants include squeezed-state, coherent-state, and two-way protocols.

\begin{itemize}
    \item {\texttt{Squeezed-State Protocols}}: One quadrature variance is reduced at the expense of the other, tightening Eve’s estimation capability and improving robustness against certain strategies; practical use is limited by specialized optics \cite{sharma2019survey}.
    \item {\texttt{Coherent State Protocols}} Alice modulates amplitude/phase of coherent states; Bob measures with homodyne or heterodyne. Security relies on fundamental uncertainty constraints that expose intrusive measurements \cite{Djordjevic2019}.
    \item {\texttt{Two-Way Protocols}}: Bob first sends a weak probe that Alice modulates and returns. The round trip raises the bar for an attacker who must compromise both directions without detection \cite{Djordjevic2019}.
\end{itemize}

\subsection{Measurement Device-Independent QKD (MDI-QKD)}

Measurement-Device-Independent QKD (MDI-QKD) is a secure quantum key distribution protocol that removes all side-channel attacks related to measurement devices \cite{lu2018recurrent}. In MDI-QKD, Alice and Bob each send quantum signals to an untrusted relay, called Charlie, who performs Hong-Ou-Mandel (HOM) interference and announces the results. Based on this, Alice and Bob can generate a shared secret key. To make MDI-QKD more practical, researchers have introduced methods like decoy states, finite-key analysis, and reference-frame-independent techniques. MDI-QKD works well in a star-shaped network where many users connect to Charlie, but its key rate drops quickly if the distances to Charlie are not equal. To solve this, new protocol versions and optimization methods have been developed to keep performance high even in asymmetric networks \cite{lu2018recurrent}.

\subsection{Twin-Field QKD (TF-QKD)}
Twin-Field QKD (TF-QKD) retains detector-side security of measurement-device-independent QKD while breaking the repeaterless rate–distance bound to achieve much longer secure transmission distances under realistic loss conditions \cite{wang2019machine}. In TF-QKD, Alice and Bob each send phase-randomized weak coherent pulses to an untrusted central relay that performs single-photon interference measurements; by using decoy-state analysis to bound multi-photon terms even in finite-size datasets, they can derive a secure key against general attacks. Since its proposal in 2018, various TF-QKD variants have been developed to improve tolerance to misalignment errors and statistical fluctuations without sacrificing measurement-device independence \cite{dong2022optimization}.

\section{Taxonomy}
\label{sec:methods}

\begin{figure*}[t]
  \centering
  \includegraphics[width=\linewidth]{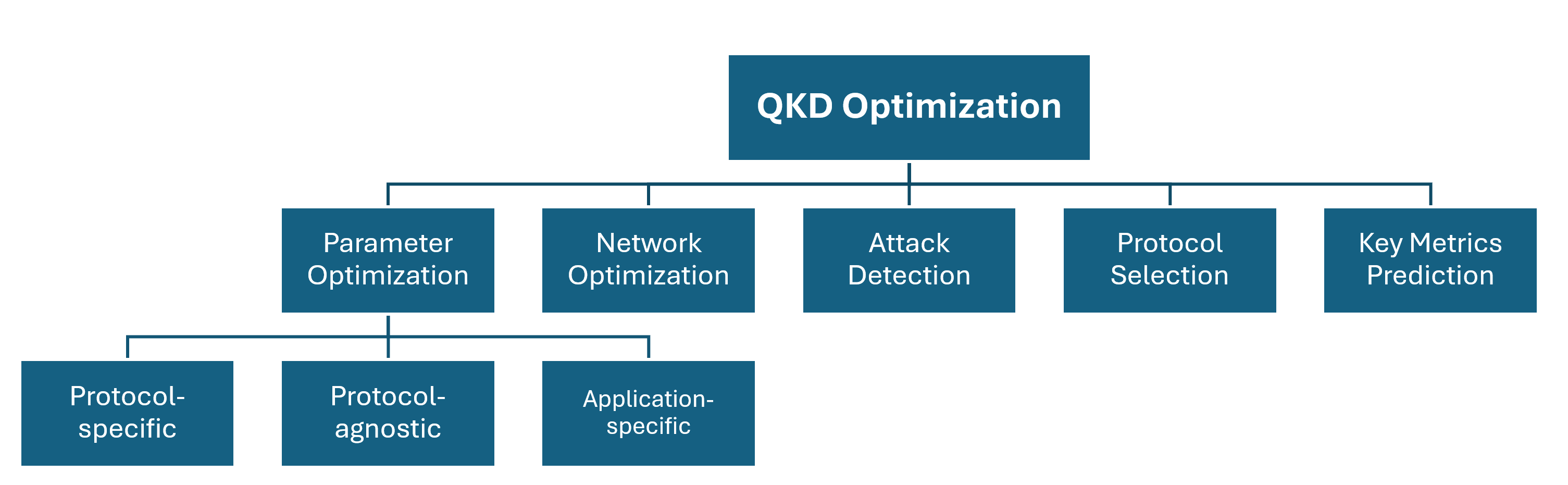}
  \caption{Taxonomy of ML methods for improving QKD}
    \label{taxonomy}

\end{figure*}

In this paper, we provide a thorough survey that systematically reviews and analyzes the existing literature by exploring how Machine Learning (ML) methods are used to improve QKD performance and security. To provide a structured approach, we categorize the areas in which ML has been used to improve QKD into five main broad categories as shown in Figure \ref{taxonomy}. 

\begin{itemize}
    \item {\texttt{Parameter optimization}}: this includes improving physical-layer and protocol-level parameters through ML-driven techniques for signal optimization, polarization drift compensation, phase stabilization, post-processing (e.g., noise suppression and error correction), and modulation control. 

    \item {\texttt{Network optimization}}: this includes the use of ML techniques in large-scale QKD networks for intelligent routing, resource allocation, and network optimization, reflecting the shift from isolated links to scalable quantum infrastructures.

    \item {\texttt{Attack detection}}: this includes the use of ML techniques to detect and classify attacks, enhancing real-time threat detection and system defense. 

    \item {\texttt{Protocol selection}}: this includes the use of ML techniques to dynamically choose or adapt QKD protocols based on environmental conditions or system feedback, thereby improving adaptability and robustness. 

    \item {\texttt{Key metrics prediction}}: this includes the use of ML techniques to predict SKRs, quantum bit error rates (QBER), and other performance indicators, enabling proactive system adjustments and more reliable operation. 
\end{itemize}

This categorization provides a comprehensive framework for understanding how ML contributes to both the physical optimization and strategic intelligence of QKD systems, highlighting its crucial role in the evolution of secure quantum communication. 
However, we note the close relationship between optimization and security. In QKD, parameter optimization is not merely a performance enhancement strategy but can also be a foundational aspect of system security. As we show in the rest of this survey, the precise tuning of parameters such as signal intensities, decoy state probabilities, polarization alignment, and modulation states directly impacts the system’s ability to limit information leakage, detect quantum attacks, and maintain high secure key generation rates under realistic channel and hardware conditions. 

\section{ML for QKD Parameter Optimization}

Many studies contributed significantly to the field of QKD by addressing one of the primary challenges: parameter optimization. The key drawback of utilizing conventional methods in parameter optimization lies in their computational inefficiency and resource-intensive nature, which often results in prohibitive delays or the reliance on sub-optimal configurations during real-time operation. Parameter optimization is the systematic selection of optimal system parameters (e.g., equipment or measurement parameters) \cite{liu2017monitoring}, physical parameters of signals (e.g., the intensity and their corresponding probabilities of transmission, phase, and polarization of the laser signal) \cite{liu2018integrating}, and the raw secret key \cite{li2018discrete}. It becomes an even more challenging problem in scenarios where QKD is deployed in multi-node networks requiring optimization. In particular, statistical fluctuations due to a finite-size key can drastically affect the performance of a practical communication system \cite{mafu2024advances}. 

For example, the authors in \cite{bommi2023enhancing}  proposed a comprehensive multi-layered architecture that integrates quantum components with advanced ML techniques to enhance the efficiency and resilience of QKD systems. The model comprises three key elements: a quantum subsystem, an ML subsystem, and a hybrid interface that facilitates real-time communication between them. The quantum subsystem includes standard QKD components, such as photon sources, detectors, and quantum channels, responsible for generating and transmitting secure keys.
The ML subsystem introduces three complementary learning strategies tailored to specific layers of the QKD process. A DL model is employed to optimize quantum state preparation, specifically targeting photon polarization alignment, thereby improving the key generation rate by 25\%. An unsupervised learning model is used for anomaly detection at the detection layer, increasing the eavesdropping detection rate by 7.8\%. Finally, a RL agent adapts system parameters, such as modulation settings and encoding basis, to environmental changes, reducing the QBER by 33.3\%. These components are integrated via a hybrid interface that converts quantum state data into ML-compatible formats, and vice versa, enabling continuous feedback and dynamic reconfiguration. The result is a robust and adaptive QKD framework that eliminates the need for manual recalibration, increases system uptime, and yields an overall system efficiency improvement of 8.2\%.

In this survey, we categorize the optimization techniques into three main categories: protocol-specific optimization, application-specific optimization, and protocol-agnostic optimization.
\subsection{Protocol-specific Optimization}

Protocol-specific optimization attempts to enhance the physical parameters of the signals in QKD, which involves selecting the best light intensities and the probabilities of sending these intensities to maximize the SKR, which is essential due to the limited number of signals that can be sent within the finite duration of QKD experiments. To illustrate the range of approaches in protocol-specific optimization, Table \ref{tab:merged_qkd_ml_comparison} provides a consolidated comparison of representative techniques and their reported performance gains in QKD systems.

\begin{table*}
  \caption{Comparison of ML-based approaches for protocol-specific optimization in QKD systems}
  \label{tab:merged_qkd_ml_comparison}
  \centering
  \small
  \renewcommand{\arraystretch}{1.2}
  \begin{tabular}{@{}p{0.09\linewidth} p{0.1\linewidth} p{0.15\linewidth} p{0.12\linewidth} p{0.13\linewidth} p{0.13\linewidth} p{0.12\linewidth}@{}}
    \toprule
    \textbf{Paper} & \textbf{Protocol type} & \textbf{Optimization type} & \textbf{ML model} & \textbf{Performance} & \textbf{Advantages} & \textbf{Limitations} \\
    \midrule
    \cite{lu2019parameter} 
      & Asymmetric MDI-QKD (3-intensity) 
      & Calibration-aware optimization (intensities, probabilities, misalignments $\Delta\phi$, $e_d$) 
      & BPNN 
      & $100\times$ speedup vs.\ LSA; prediction time $\sim 0.186$\,s; key rate $\approx 85\%$ of optimal 
      & Real-time hardware calibration (phase and polarization compensation) 
      & Protocol-specific; lower key rate than optimal \\
    \addlinespace[1pt]
    \cite{wang2019machine} 
      & 4-/7-intensity MDI-QKD, BB84, TF-QKD 
      & Generalized optimization of intensities and probabilities (6–12 params by protocol) 
      & FCNN (2 hidden layers, ReLU) 
      & $10^2$–$10^4\times$ speedup; inference 1–3\,ms; key rate 95–99.99\% 
      & Universal framework across protocols; efficient on embedded devices 
      & Assumes fixed hardware; no real-time calibration \\
    \addlinespace[1pt]
    \cite{cao2025comparative} 
      & 4-intensity MDI-QKD 
      & Fast prediction of optimal intensities ($\mu,\nu,\omega$), transmission and basis probabilities 
      & RF, MLP, GLM, XGBoost, KNN, Decision Tree 
      & Inference in $\mu$s; $R^2 \approx 0.99$; key rate 97–99\% of optimal 
      & Very fast; high accuracy; multiple ML baselines compared 
      & Protocol-specific; no calibration features \\
    \addlinespace[1pt]
    \cite{liu2018integrating} 
      & GMCS-CVQKD 
      & Stabilize LO/signal intensity drift (not protocol parameters) 
      & SVR 
      & Improved LO stability; reduced excess noise; maintained phase reference 
      & Enhances CV-QKD reliability; deployable in real time 
      & Not for protocol parameter optimization; limited to signal stabilization \\
    \addlinespace[1pt]
    \cite{ding2020predicting} 
      & 3-intensity MDI-QKD and BB84 
      & Predict optimal protocol parameters (intensities, probabilities, bases) 
      & RF regression 
      & Accuracy $>99\%$; key rate 99.41\% of optimal; 2\,s for 8000 predictions 
      & High accuracy; fast; protocol-agnostic; scalable 
      & Needs large labeled sets; depends on simulation fidelity \\
    \addlinespace[1pt]
    \cite{yi2021optimization} 
      & 3-intensity MDI-QKD 
      & Refined RF optimization with pruning/normalization 
      & Enhanced RF 
      & $R^2 = 0.995$, MSE $= 1.81\times10^{-4}$; key rate near optimal 
      & Better generalization vs.\ vanilla RF 
      & No experimental comparison; large dataset required \\
    \addlinespace[1pt]
    \cite{kang2023machine} 
      & Symmetric or asymmetric 3-intensity TF-QKD 
      & Predict optimal $s,\mu_0,P_X,P_{\mu_0}^Z$ (incl.\ asymmetric extensions) 
      & BPNN, RBFNN, GRNN 
      & Key rate $\approx 99.97\%$ of LSA (RBFNN); MSE $10^{-9}$–$10^{-5}$; 4.3–27\,$\mu$s 
      & High accuracy; real-time; supports asymmetry 
      & No feature importance; trained on LSA-generated data \\
    \addlinespace[1pt]
    \cite{dong2022optimization} 
      & Symmetric 3-intensity TF-QKD 
      & Predict optimal $\mu,\nu$ 
      & XGBoost, RF, BPNN 
      & Key-rate ratio 99.63\%; MSE $\sim 2.2\times10^{-5}$; 0.005\,s (3000 samples) 
      & Fastest; lowest MSE; interpretable feature importance 
      & Limited to $\mu,\nu$; no asymmetric modeling \\
    \bottomrule
  \end{tabular}
\end{table*}

The first foundational work by the authors in~\cite{lu2019parameter} established the viability of NNs for real-time calibration in MDI-QKD. Their backpropagation neural network (BPNN) simultaneously predicts optimal decoy-state parameters (\textit{$\mu$, $\nu$}, basis probabilities) and compensates for phase/polarization misalignments using discarded communication data. This approach achieves remarkable speed (0.186\,s inference) while maintaining 80--100\% of optimal key rates, though its MDI-QKD specificity limits cross-protocol applicability.

Building on this, the authors in \cite{wang2019machine} developed a protocol based on a unified neural network architecture. Their fully-connected NN with ReLU activations enables multi-protocol optimization (BB84 to TF-QKD) without retraining, achieving 100-1000  speed improvements over conventional methods while preserving 95-99.99\% of maximum key rates. 

Recent work \cite{cao2025comparative} addresses temporal constraints in four-intensity MDI-QKD through a novel double-scanning pipeline. The comparative metrics in their approach combine coarse worst-case scanning with ML-optimized fine-scanning, enabling microsecond-scale inference. Benchmarking six regressors revealed RF's superior accuracy, $R^2$ equals 0.99, and MLP's robustness under channel fluctuations outperforming earlier approaches in dynamic channel conditions.

Complementing these discrete-variable approaches, the authors in \cite{liu2018integrating} pioneered ML stabilization for CV-QKD using support vector regression (SVR). Their method achieves theoretical optimal key rates without additional hardware. Unlike the preceding decoy-state optimizations, their method predicts laser intensity/Local Oscillator (LO) drifts for real-time feedback control. The SVR's lightweight computation (1.2\,ms/cycle) demonstrates ML's versatility across QKD paradigms, from parameter prediction to physical layer stabilization.

Two notable contributions have demonstrated the effectiveness of RF regression for predicting optimal QKD parameters. Ding \emph{et al.}~\cite{ding2020predicting} and Yi \emph{et al.}~\cite{yi2021optimization} both employed RF models to optimize 3-intensity decoy-state MDI-QKD systems, targeting key parameters such as signal/decoy intensities ($\mu$, $\nu$), their probabilities ($P_\mu$, $P_\nu$), and basis selection probabilities ($P_{X|\mu}$, $P_{X|\nu}$). However, their methodologies and scopes differ significantly.

\begin{itemize}
    \item Ding \emph{et al.} \cite{ding2020predicting} adopted a standard RF model and rigorously benchmarked it against traditional optimization methods (LSA) and neural networks (NNs). Their results showed that RF achieves $>99\%$ prediction accuracy and nearly matches the optimal SKR obtained via LSA, while reducing inference time dramatically, processing 8,000 parameter sets in just 2 seconds. This makes their approach particularly suitable for real-time, large-scale QKD networks. Notably, they extended their model to support both MDI-QKD and BB84 protocols, demonstrating broader applicability.
    \item Yi \emph{et al.} \cite{yi2021optimization} proposed an enhanced RF model incorporating data preprocessing and tree pruning based on Out-of-Bag (OOB) testing. Their refined algorithm achieved strong performance metrics ($R^2 = 0.995$, mean squared error (MSE) = 0.000181), suggesting high predictive accuracy. However, their study focused exclusively on MDI-QKD and did not experimentally compare their enhanced RF to other ML baselines (e.g., standard RF or NNs). While their approach claims improved generalization, \textit{real-time capability} its practical advantages remain conceptual without empirical validation against alternative methods.
    
\end{itemize}

ML for TF-QKD parameter optimization are exemplified by two representative studies.
\begin{itemize}
    \item Kang \emph{et al.}~\cite{kang2023machine} implemented a comprehensive neural network framework comprising BPNN, Radial Basis Function Neural Network (RBFNN), and General Regression Neural Network (GRNN) architectures for both symmetric and asymmetric TF-QKD configurations. Their results demonstrated that RBFNN achieved the highest fidelity in parameter prediction, maintaining key rates within 99.97\% of LSA benchmarks, while BPNN exhibited superior computational efficiency with inference times on the order of microseconds.
    \item Dong \emph{et al.}~\cite{dong2022optimization} adopted an alternative approach using XGBoost - a gradient-boosted tree ensemble method - in conjunction with BPNN and RF models. Their methodology focused specifically on symmetric TF-QKD implementations, optimizing exclusively the signal intensity ($\mu$) and decoy intensity ($\nu$) parameters. The XGBoost implementation not only achieved a key rate ratio of 99.63\% relative to exhaustive search results with minimal MSE, but also provided valuable feature importance analysis, identifying transmission distance as the predominant optimization factor. Notably, their framework demonstrated a 300,000$\times$ acceleration in prediction time compared to conventional search algorithms.
\end{itemize}

Despite both studies employing supervised learning for parameter prediction in TF-QKD, they differ notably in scope and methodology. Kang \emph{et al.} prioritize broader protocol coverage and leverage the flexibility of neural networks to model both symmetric and asymmetric scenarios. In contrast, Dong \emph{et al.} focus on computational efficiency and interpretability by utilizing tree-based models, particularly XGBoost. While neither study explicitly addresses the generalizability of their approach to other QKD protocols, the underlying methods are adaptable and could be extended with appropriate retraining. Overall, the two approaches exhibit complementary strengths, making them suitable for distinct deployment contexts in real-time QKD systems.

\subsection{Protocol-agnostic Optimization}
Unlike protocol-specific optimization techniques, protocol-agnostic optimization focuses on controlling phase, polarization, post-processing, and modulation variance to enhance both the practical security and key-rate performance of QKD systems.
The comparative overview in Table~\ref{tab:unified_qkd_ml_comparison} highlights how ML contributes to different stages of QKD. While polarization optimization ensures stable transmission, phase stability optimization addresses coherence over long distances, and post-processing optimization enhances reconciliation efficiency. Together, these approaches underline the versatility of ML for improving both physical-layer stability and protocol-layer performance in QKD systems.

\begin{table*}
  \caption{Comparison of ML-based approaches for protocol-agnostic optimization in QKD systems}
  \label{tab:unified_qkd_ml_comparison}
  \centering
  \footnotesize
  \renewcommand{\arraystretch}{1.2}
  \begin{tabular}{@{}p{0.08\linewidth} p{0.1\linewidth} p{0.13\linewidth} p{0.13\linewidth} p{0.14\linewidth} p{0.13\linewidth} p{0.13\linewidth}@{}}
    \toprule
    \textbf{Paper} & \textbf{Protocol type} & \textbf{Optimization type} & \textbf{ML model} & \textbf{Performance} & \textbf{Advantages} & \textbf{Limitations} \\
    \midrule
    \cite{zhou2024adjusting}
      & CV-QKD
      & Polarization optimization
      & ML-based dynamic polarization control (real-time)
      & Improved SKR; reduced QBER; mitigated polarization distortions
      & Enhances polarization stability; protects against polarization-based attacks
      & Higher compute; precise optics needed; may degrade under unseen turbulence \\
    \cite{ahmadian2022cost}
      & DV-QKD (BB84)
      & Polarization optimization
      & DNN + polynomial fitting for SOP prediction (real-time)
      & SKR $\sim$4.5\,Mb/s; QBER from 6.1\% to $>1.5\%$
      & Fewer false alarms; maintains secure exchange under stress
      & Large training sets; costly SOP tracking; weak under extreme stress \\
    \addlinespace[2pt]
    \cite{chin2021machine}
      & CV-QKD
      & Phase stability optimization
      & UKF
      & Lower excess noise; stable carrier-phase recovery
      & Outperforms EKF; robust phase tracking
      & Sensitive to pilot noise; fine-tuning required \\
    \cite{liu2019practical}
      & DV-QKD (BB84, phase encoding)
      & Phase stability optimization
      & LSTM
      & Duty cycle from 50\% to 83\%; faster calibration with similar accuracy
      & Matches accuracy with less calibration time
      & Periodic retraining; updated datasets needed \\
    \cite{zhang2021machine}
      & MDI-QKD
      & Phase stability optimization
      & LSTM
      & Better phase alignment; fewer external calibrations
      & No stabilization lasers; supports long-distance links
      & Dependent on training quality; retraining required \\
    \cite{hajomer2024long}
      & CV-QKD
      & Phase stability optimization
      & UKF
      & Enables $100$\,km LLO CV-QKD with finite-size security
      & Practical long distance without complex stabilization
      & Careful tuning; $V_{\mathrm{mod}}$ optimized experimentally (not by ML) \\
    \cite{long2024phase}
      & CV-QKD (satellite-to-ground)
      & Phase stability optimization
      & CNN
      & Higher coherent efficiency; SKR without wavefront sensors
      & No direct phase sensors; simpler design
      & Accuracy limited by CNN generalization; simulation-trained models \\
    \addlinespace[2pt]
    \cite{chen2025machine}
      & CV-QKD (satellite-to-ground, Gaussian-modulated)
      & Modulation variance optimization
      & ResFCN18 NN
      & $\sim$99.5\% $V_{\mathrm{mod}}$ prediction accuracy; $10^{3}$–$10^{4}$× faster compute; SKR gain up to 22\,dB; secure window +192\,s at 710\,km
      & Near real-time $V_{\mathrm{mod}}$ tuning on LEO links; low-power feasible; resilient to fast fluctuations
      & Needs large simulated orbital sets; generalization tied to turbulence assumptions \\
    \cite{jiangmachine}
      & CV-QKD (fiber, Gaussian-modulated)
      & Modulation variance optimization
      & GA–BP NN
      & Converges to globally optimized $V_{\mathrm{mod}}$; steadier SKR across lengths/noise
      & GA global search + BP convergence; robust to fiber fluctuations
      & Higher compute than direct predictors; slower than CNN/ResFCN; less suited to short satellite windows \\
    \addlinespace[2pt]
    \cite{sihare2024analyzing}
      & QKD (error-correction focus)
      & Post-processing optimization
      & Error prediction + stabilizer codes
      & QBER $\downarrow$ 12\%; SKR $\uparrow$ 15\%
      & More efficient post-processing
      & Computationally intensive \\
    \cite{liang2022machine}
      & CV-QKD
      & Post-processing optimization
      & NN + KNN classification
      & Excess noise $\downarrow$ 72\%; error variance $\downarrow$ 26\%; SKR $\uparrow$ 18\%
      & Strong noise reduction; better reconciliation
      & Requires accurate pilot calibration \\
    \cite{zhang2022neural}
      & CV-QKD
      & Post-processing optimization
      & LSTM autoencoder (nonlinear compensation)
      & 9.57\,Mbps (500\,MHz) and 18\,Mbps (1\,GHz) SKR; excess noise $\sim10^{-3}$ SNU
      & High-speed CV-QKD under nonlinear effects
      & Needs pre-trained deep models; limited adaptability to dynamics \\
    \bottomrule
  \end{tabular}
\end{table*}

\subsubsection{Polarization Optimization}
Polarization optimization in QKD addresses the random fluctuations in the state of polarization (SOP) caused by environmental disturbances. These fluctuations can increase QBER and risk information leakage if not managed properly. Conventional reactive compensation methods often introduce operational overhead and fail to adapt quickly under severe SOP variations. To address these limitations, ML has been proposed to predict SOP drift and proactively adjust polarization control mechanisms in real time, even under challenging conditions. 
Several studies expanded on this by optimizing polarization fluctuations, providing a more nuanced understanding of the challenge with atmospheric turbulence affecting CV-QKD  that focused on dynamic polarization shifts in free-space channels \cite{zhou2024adjusting} and mechanical fiber stress disrupting BB84 QKD, predicting SOP drift in optical fibers \cite{ahmadian2022cost}. 
They used real-time ML models to predict polarization state changes. In both cases, ML-based prediction enables proactive polarization correction strategies, dynamic polarization control in CV-QKD and electronic polarization controllers in BB84, reducing polarization misalignment and QBER. As a result, both methods enhanced SKR stability and improved eavesdropping detection robustness.

\subsubsection{Phase Stability Optimization}
Phase stability is maintaining consistent and predictable phase relationships between the transmitted and received quantum signals during key distribution. In phase-encoded or coherent detection QKD systems, the quantum states' phase can drift due to environmental disturbances (e.g., temperature, humidity, mechanical vibrations) and internal device imperfections (e.g., laser phase noise, interferometer drift). This phase drift leads to instability in interference conditions at the receiver and affects key performance metrics of the QKD system. Phase stability optimization involves actively controlling and compensating for phase fluctuations. Techniques include real-time calibration, which is essential because frequent phase drift degrades system stability and lowers SKR, especially in practical, high-speed QKD networks. Real-time calibration enables continuous compensation of phase fluctuations without interrupting key transmission, thus improving both stability and throughput.

The optimization of phase stability in QKD systems has been the focus of several recent works, where real-time ML methods have been employed to predict and correct phase noise. Chin \emph{et al.}~\cite{chin2021machine} implemented an Unscented Kalman Filter (UKF) for real-time carrier phase tracking in CV-QKD, effectively mitigating excess noise from laser drift and frequency instability while ensuring accurate phase estimation. Liu \emph{et al.}~\cite{liu2019practical} utilized an LSTM model in DV-QKD to predict and correct phase modulation drift in real-time, overcoming inefficiencies of traditional scanning-and-transmitting calibration methods and improving the system's duty cycle from 50\% to 83\% without compromising QBER performance. Similarly, Zhang \emph{et al.}~\cite{zhang2021machine} addressed reference frame misalignment in MDI-QKD using LSTM-based real-time phase drift prediction, significantly reducing the need for external stabilization lasers. Expanding on these approaches, Hajomer \emph{et al.}~\cite{hajomer2024long} employed a ML-aided UKF framework for carrier phase recovery in CV-QKD, stabilizing phase against laser noise in a practical 100~km fiber implementation while maintaining high SKR, and additionally optimized the system's modulation variance experimentally to further enhance performance. Long \emph{et al.}~\cite{long2024phase} applied a CNN for real-time phase wavefront correction using intensity-only measurements in satellite-to-ground CV-QKD, improving coherent efficiencies and enabling positive key rates without complex phase-sensing hardware. 

In all of these studies, real-time ML enabled prediction and compensation allowed for proactive phase correction strategies, replacing or reducing the frequency of reactive recalibration steps, thus improving SKR stability and overall system robustness.

\subsubsection{Modulation Variance Optimization}
Modulation Variance (Vmod) means the variance of the Gaussian distribution used to modulate the amplitude and phase quadratures of coherent optical states, determining the signal’s strength and directly impacting the mutual information, excess noise contribution, and the achievable SKR of the system.
The authors in \cite{chen2025machine} addressed the challenge of fast and efficient Vmod optimization in free-space Gaussian-modulated CV-QKD, particularly under the strict power and computational constraints of satellite-to-ground links. Recognizing that conventional LSA for determining the optimal Vmod is too slow for the short access windows of low-Earth-orbit satellites, the authors proposed a ResFCN18 NN framework capable of predicting the optimal Vmod in near real-time. By employing a simulation platform that integrates precise orbital models to generate realistic training data, the model learns to map environmental and system conditions to the Vmod that maximizes SKR under varying channel conditions. Their results demonstrated an accuracy of approximately 99.5\% in predicting the optimal vmod while reducing computation time by 3--4 orders of magnitude, enabling rapid, adaptive parameter optimization essential to secure global quantum communication networks. Similarly, Jiang \emph{et al.}~\cite{jiangmachine} proposed a GA-BP NN framework for Vmod optimization to the same QKD protocol, combining a genetic algorithm for global search with a backpropagation network for efficient convergence to maximize SKR under practical channel conditions. While Chen \emph{et al.} leveraged fast, direct prediction for dynamic free-space channels, Jiang \emph{et al.} focused on robust, globally optimized Vmod tuning under fiber-based conditions, demonstrating the complementary roles of DL prediction and hybrid search-optimization frameworks in enhancing practical CV-QKD systems.

\subsubsection{Post-Processing Optimization}
Several studies leverage ML to optimize parameters within the post-processing regime of QKD systems, aiming to enhance the extraction of secure keys. Liang \emph{et al.} \cite{liang2022machine} propose an ML-assisted equalization scheme designed to suppress excess noise prior to the reconciliation stage, effectively optimizing the signal conditioning process that precedes traditional post-processing. This approach seeks to reduce the burden on subsequent reconciliation steps. Sihare \cite{sihare2024analyzing} explored the optimization of Quantum Error Correction (QEC) strategies, a crucial post-processing stage, through ML-based error prediction. By predicting and correcting errors more efficiently, the parameters of the QEC process itself are optimized. Zhang \emph{et al.} \cite{zhang2022neural} introduce an autoencoder-based NN framework to compensate for nonlinear distortions in CV-QKD. This framework optimizes the accuracy of the quadrature measurements at the receiver, a post-processing step that is essential for accurate parameter estimation and secure key extraction.

\subsection{Application-specific Optimization}

ML has also been applied to optimize QKD performance in application-specific scenarios, particularly in underwater environments where optical channels are strongly degraded by turbulence, scattering, and absorption. Unlike general-purpose optimization strategies, these studies target concrete physical bottlenecks within distinct QKD implementations, demonstrating how ML can be customized to maximize system performance under environmental and hardware constraints.   

Two recent works illustrate complementary directions for such application-specific optimization~\cite{nozari2024deep,yi2024passive}. The first focuses on the detection side of a discrete-variable BB84 system, while the second centers on the channel side of a (CV-MDI) QKD protocol.   

\begin{itemize}
    \item {Nozari and Uysal} \cite{nozari2024deep} investigated a BB84 system with single-photon avalanche diodes (SPADs), using a DNN trained with the Adam optimizer to predict optimal detector parameters. Specifically, their model jointly optimized bit time, field of view, and gate time to minimize QBER caused by background noise and intersymbol interference. The approach achieved a mean squared error (MSE) on the order of $10^{-4}$ with $R^2$ values approaching unity, indicating highly accurate predictions. When applied to the QKD system, this resulted in QBER suppression by up to two orders of magnitude across 10--40 m underwater links, underscoring the importance of precise detector control in noisy optical channels.  

    \item {Yi} \emph{et al.} \cite{yi2024passive} focused instead on channel characteristics by developing an ML model for passive (CV-MDI) QKD.  They proposed a hybrid GA-Elman recurrent neural network to predict transmittance as a function of depth and distance, thereby reducing the need to sacrifice raw keys for parameter estimation. Their method improved prediction accuracy substantially, lowering average error from 2.814\% with a standard Elman network to 0.506\% with GA-Elman. This accuracy translated directly into higher SKR, since more reliable transmittance estimation enabled more efficient key generation even under strong optical fading. 
\end{itemize}
 
Taken together, these studies demonstrate how ML can be tailored to distinct components of underwater QKD: detector optimization in discrete-variable protocols and channel prediction in continuous-variable protocols. While both approaches achieved significant performance gains, they also face constraints: SPAD-based optimization is limited to prepare-and-measure schemes like BB84, whereas transmittance prediction is bounded by the attenuation properties of seawater.

\section{ML for QKD Networks Optimization}

ML has been used extensively in quantum networks to improve the security of QKD protocols by enhancing the detection of possible attacks and optimizing the key rate. Table~\ref{tab:ml_qkd_networks} provides a concise comparison between three recent related studies. We provide more details on these studies below.

Quantum-classical coexistence in Multi-Band Elastic Optical Networks (EONs) is one of the biggest challenges in quantum networks, where quantum channels have to share the same fiber infrastructure with high-power classical channels. These limit the performance of quantum communication due to the intense noise and interference caused by the classical channels, affecting the SKR and QBER. In this case, ML-based approaches can be used to optimize the frequency in the quantum channel. For example, in \cite{mehdizadeh2024ml}, the authors demonstrated that ML algorithms, such as NN, KNN, and XGB, can achieve an accuracy of 99\% in predicting the best quantum channel frequency in real time, thus allowing the reduction of the computation time from 637 to 0.09 seconds.

Similarly, the authors in \cite{ahsun2025adaptive} proposed an ML-driven framework that dynamically tunes QKD parameters such as photon transmission rate, QBER thresholds, and error correction schemes based on real-time network noise. The model was demonstrated to be effective across different deployment settings, including fiber, satellite, and high-speed QKD deployment, reflecting strong generalizability. Their approach uses supervised learning, reinforcement learning, and genetic algorithms to achieve improved SKR and reduced QBER. 

Finally, the authors in \cite{johann2023routing} proposed an LSTM-based traffic prediction model that predicts future traffic matrices due to limited key generation rates and inefficient routing that can lead to depletion of the keystore, causing delays or failures in secure communication. The model allows the network to dynamically adjust the routing weights before a keystore is exhausted. The model was evaluated against both a hop-count shortest-path algorithm and a hypothetical perfect prediction model. The results show a significant reduction in blocking probability while avoiding excessive information sharing.

\begin{table*}[H]
  \caption{ML-driven approaches for enhancing QKD security in quantum communication networks}
  \label{tab:ml_qkd_networks}
  \centering
  \footnotesize
  \renewcommand{\arraystretch}{1.2}
  \begin{tabular}{@{}p{0.11\linewidth} p{0.1\linewidth} p{0.14\linewidth} p{0.10\linewidth} p{0.10\linewidth} p{0.10\linewidth} p{0.10\linewidth} p{0.10\linewidth}@{}}
    \toprule
    \textbf{Paper} & \textbf{ML model} & \textbf{Application} & \textbf{SKR} & \textbf{QBER} & \textbf{Efficiency} & \textbf{Detection} & \textbf{Limitations} \\
    \midrule
    \cite{mehdizadeh2024ml}
      & KNN, NN, XGB
      & Optimize quantum–classical coexistence in multi-band EONs to maximize SKR and reduce classical noise
      & Accuracy 99\% for optimal QCh frequency; compute $637$\,s $\rightarrow$ $0.09$\,s
      & Indirectly improved by lowering classical-noise interference
      & Real-time processing (large speedup)
      & Not the focus (performance over attack prevention)
      & Narrow focus on frequency; limited beyond coexistence scenarios \\
    \cite{ahsun2025adaptive}
      & Supervised + RL; genetic algorithms; NNs
      & Adaptive tuning of QKD parameters (e.g., photon rate, QBER) under noise
      & Improves SKR in fiber, satellite, and high-speed settings via noise-aware adaptation
      & Reduced via dynamic adjustment of transmission parameters
      & Real-time optimization via predictive modeling
      & Emphasis on error reduction, not specific threat detection
      & Scalability challenges; relies on accurate noise models \\
    \cite{johann2023routing}
      & LSTM
      & Predictive routing to avoid keystore exhaustion in meshed QKD networks
      & Prevents blocking by rerouting before key depletion; stable up to traffic factor 500
      & Balanced link usage indirectly lowers QBER
      & Fast, preemptive routing decisions
      & Not addressed (focus on reliability/resources)
      & Depends on historical traffic; accuracy drops at extreme load \\
    \bottomrule
  \end{tabular}
\end{table*}

\section{ML for QKD Attack Detection}

Recent work has turned to ML as a practical way to detect attacks in QKD. Instead of fixed threshold rules, models are trained from data to learn normal behavior and flag deviations. Both supervised and unsupervised settings appear in the literature, operating on core physical-layer signals such as quadrature records, LO power, and noise-related measures. With datasets drawn from simulations and experiments, these methods handle a wide span of threats—calibration and LO manipulation, saturation/wavelength effects, Trojan horse, impersonation, photon splitting, DoS, and hybrids—and can also surface patterns that were not explicitly modeled.

Evidence spans CV-QKD and DV-QKD. Table~\ref{tab:cvqkd_and_dvqkd_ml_attacks} groups representative studies and contrasts their data sources, attack coverage, model choices, and trade-offs. In CV-QKD, Du and Huang~\cite{du2022multi} treat the task as multi-label and use BR-NN/LP-NN with a one-class SVM stage to address out-of-distribution samples. Kish \emph{et al.}~\cite{kish2024mitigation} target channel amplification and related DoS; a lightweight decision tree detects attack states and a post-selection step improves SKR when the channel is tampered. Mao \emph{et al.}~\cite{mao2020detecting} adopt a feature-based ANN built on compact descriptors (quadratures, LO level, noise variance) to separate normal from multiple attack/hybrid cases using simulated data. Luo \emph{et al.}~\cite{luo2022beyond} propose a semi-supervised deep approach that learns a universal statistical model from physical-layer measurements and generalizes to known and previously unseen conditions. Complementing these, Ding \emph{et al.} \cite{ding2023machine} introduced a universal ML pipeline (MADS) that constructs block‐level CV-QKD feature vectors, applies DBSCAN for noise/outlier pruning, and then performs multi-class classification using a one-vs-rest SVM with a Gaussian kernel. Using an eight-class dataset (no-attack; LO-intensity, calibration, saturation, wavelength, blinding; two hybrids), the authors report near-perfect detection performance (accuracy/precision/recall/F1 $\approx 0.9996$) and demonstrate that MADS tightens overestimated SKR, yielding safer key-rate bounds. The method is fast and lightweight and exhibits strong robustness to outliers. More recently, Jin \emph{et al.} \cite{jin2024attack} modeled CV-QKD monitoring data as graphs and used a graph isomorphism network (GIN) with three graph-building methods: Method I forms a multi-parameter graph from four monitored features—shot-noise variance, local-oscillator power, electronic noise, and quadrature variance; Method II builds a graph from the parameter-estimation pairs and the block-wise channel-transmittance ratio (no additional monitoring hardware); and Method III combines I+II by embedding both structure and feature labels. Training in data at $30,\mathrm{km}$ with modulation variance $V_A{=}2$ and tested in five independent sets ($30$–$70,\mathrm{km}$; $V_A{\in}{2,4,6}$), their scheme achieves $97.8\%$ (Method I), $86.4\%$ with $100\%$ attack vs. no-attack detection efficiency (Method II), and $98.8\%$ (Method III) on the in-distribution test; Methods I/III maintain $>95\%$ accuracy across transfer sets (distance and $V_A$ changes). Also, their anomaly-detection module on Method III—flag partially unknown attacks with $\sim85\%$ effectiveness.

\begin{table*}[H]
  \caption{Comparison of ML-based attack detection in CV-QKD and DV-QKD systems}
  \label{tab:cvqkd_and_dvqkd_ml_attacks}
  \centering
  \footnotesize
  \renewcommand{\arraystretch}{1.2}
  \begin{tabular}{@{}p{0.05\linewidth} p{0.06\linewidth} p{0.12\linewidth} p{0.12\linewidth} p{0.10\linewidth} p{0.10\linewidth} p{0.07\linewidth} p{0.10\linewidth} p{0.10\linewidth}@{}}
    \toprule
    \textbf{Paper} & \textbf{Type} & \textbf{Training Data} & \textbf{Attacks} & \textbf{Model} & \textbf{Accuracy} & \textbf{Gen.} & \textbf{Pros} & \textbf{Cons} \\
    \midrule
    \cite{luo2022beyond}  
      & CV 
      & Experimental CV-QKD with homodyne detection 
      & Calibration, LO intensity, saturation, wavelength, unknown 
      & GAN + anomaly detection 
      & 99.3\% (known) 
      & High 
      & Real-time detection; strong generalization 
      & High model complexity (GAN) \\

    \cite{mao2020detecting}  
      & CV 
      & Simulated CV-QKD parameters 
      & Calibration, LO intensity, saturation, hybrid 
      & ANN (feature-based) 
      & $>99\%$ precision/recall 
      & Moderate 
      & Multi-attack classification 
      & Slight SKR/transmission reduction \\

    \cite{du2022multi}     
      & CV 
      & Simulated CV-QKD parameters 
      & Calibration, LO intensity, saturation 
      & BR-NN \& LP-NN + one-class SVM 
      & 100\% (known), 99.8\%, 98.7\% 
      & High 
      & Detects known \& unknown attacks 
      & One-class SVM false positives \\

    \cite{kish2024mitigation}  
      & CV 
      & Analytical CV-QKD parameter estimates 
      & Channel amplification (CA), CA-DoS, DoS 
      & Decision tree classifier 
      & 100\% (low noise), 90.1\% (high) 
      & Low 
      & Fast, lightweight classification 
      & Narrow threat scope \\

    \cite{jin2024attack}
      & CV
      & Simulated CV-QKD parameters (distance and multiple modulation settings)
      & Calibration, LO power fluctuation, saturation, wavelength, unknown 
      & GIN (I: multi-feature graph; II: transmittance-only; III: hybrid)
      & 97.8\% (I), 86.4\% (II), 98.8\% (III); unknown $\sim$85\% (III)
      & High
      & No extra monitors (II); strong transfer (I, III); unknown-attack option (III)
      & II confuses attack \emph{types}; graph design complexity; needs block-wise data \\

    \cite{ding2023machine}
      & CV
      & Experimental CV-QKD parameters
      & LO intensity, calibration, saturation, wavelength, blinding, hybrid (CA+SA, CA+LO)
      & DBSCAN + multiclass SVM (Gaussian kernel)
      & 99.96\%
      & High
      & Lightweight; outlier pruning; improves SKR tightness
      & Requires feature engineering; kernel selection sensitivity \\

    \cite{al2021machine}   
      & DV 
      & Simulated DV-QKD key-length data 
      & Man-in-the-middle attacks 
      & ANN + LSTM 
      & 99.1\% 
      & Low 
      & Effective for predefined attacks 
      & High LSTM compute time \\

    \cite{tunc2023machine}  
      & DV 
      & Theoretical BB84 DV-QKD simulations 
      & Eavesdropping attacks 
      & SVM + LSTM 
      & 100\% precision/recall 
      & Low 
      & Accurate compromised-channel detection 
      & Limited attack variety \\

    \cite{xu2024automatically} 
      & DV 
      & Experimental + simulated imperfection data 
      & Device imperfections, eavesdropping 
      & Random Forest classifier 
      & 98\% 
      & High 
      & Real-time detection of known \& unknown threats 
      & High model complexity \\

    \cite{chou2024empirical}  
      & DV  
      & Experimental QBER time series (1 m, 1 km, 30 km)  
      & Trojan-horse pulse-injection attacks  
      & Gaussian Mixture Model + Bayesian classifier  
      & $>98\%$
      & Low  
      & Adaptive, risk-aware thresholds for long-distance detection  
      & Focused only on Trojan-horse attacks \\ 
    \bottomrule
  \end{tabular}
\end{table*}

Other DV-QKD studies focus on BB84~\cite{al2021machine,tunc2023machine,xu2024automatically}. Alkuwari \emph{et al.}~\cite{al2021machine} consider IoT deployments and run ANN/LSTM on controller-side resources to detect man-in-the-middle attacks. Tunc \emph{et al.}~\cite{tunc2023machine} use two gatekeepers (LSTM and SVM) that mark a session as safe or compromised based on accuracy and key-length indicators without interrupting traffic. Xu \emph{et al.}~\cite{xu2024automatically} build a general device-imperfection model and use RF for real-time detection of imperfections and eavesdropping, enabling feedback to suppress deviations. Chou \emph{et al.}~\cite{chou2024empirical} track QBER time series with GMMs and a Bayesian classifier and report $>98\%$ detection of Trojan-horse pulses over fibers up to 30\,km.

\section{ML for QKD Protocol Selection}

Selecting an optimal QKD protocol for a particular scenario is key in optimizing the use of QKD. Nayana JS \emph{et al.} \cite{js2023real} applied KNN, CNN, SVM, DT, and RF to predict the best QKD protocol among BB84, MDI-QKD, and TF-QKD, based on key system parameters. The RF classifier performed the best, achieving an accuracy close to 98.2\%, significantly reducing computation time from 8–10 hours to under 1 second. It used a simulation-based dataset with around 8000 records and was evaluated using confusion matrices, ROC curves, and k-fold cross-validation. Similarly, Ren \emph{et al.} \cite{ren2020implementation} also explored protocol selection using ML, comparing RF, SVM, K-NN, MNB, and CNN to determine the best QKD protocol in different scenarios, such as varying transmission distances, detector efficiencies, and error rates. RF outperformed all other tested models, including SVM, CNN, MNB, and K-NN, with a prediction accuracy exceeding 98\%, confirming its robustness and efficiency in protocol selection. The evaluation involved accuracy, ROC AUC, and confusion matrix analysis using synthetic datasets generated across various QKD configurations. However, the limitation of this work is that it focus on only three QKD protocols: BB84, MDI-QKD, and TF-QKD, leaving its generalizability to other protocols unverified. In both studies, RF emerges as the most reliable model for protocol prediction, but real-time adaptability and universality remain key challenges in applying ML to the selection of QKD protocols at scale. This limitation recurs across studies, highlighting the need for broader protocol coverage in future research.

\begin{table*}[H]
  \caption{Comparison of QKD protocol selection using ML techniques}
  \label{tab:qkd_ml_comparison}
  \centering
  \footnotesize
  \renewcommand{\arraystretch}{1.2}
  \begin{tabular}{@{}p{0.05\linewidth} p{0.14\linewidth} p{0.10\linewidth} p{0.12\linewidth} p{0.13\linewidth} p{0.10\linewidth} p{0.13\linewidth}@{}}
    \toprule
    \textbf{Paper} & \textbf{Protocols} & \textbf{Predicted parameters} & \textbf{Comm.\ model} & \textbf{ML techniques} & \textbf{Metrics \& best model} & \textbf{Notes} \\
    \midrule
    \cite{js2023real}
      & BB84, MDI-QKD, TF-QKD
      & Dark count, misalignment, detector efficiency, \#pulses, distance
      & Detector-level optimization
      & KNN, CNN, SVM, DT, RF
      & SKR, QBER, security rankings; \textbf{RF} best (98.2\%, fast)
      & RT: yes (real-time); feedback: none; data: simulated $\sim$8000 samples; eval: confusion matrix, ROC, accuracy, $k$-fold CV; limits: only three protocols; future: extend to more protocols/dynamic conditions \\

    \cite{ren2020implementation}
      & BB84, MDI-QKD, TF-QKD
      & Same five (for key-rate prediction and selection)
      & Full stack (source / encoder / detector)
      & RF, SVM, KNN, MNB, CNN
      & Key-rate, accuracy, AUC; \textbf{RF} best ($>$98\%, highest ROC AUC)
      & RT: yes; feedback: none (theoretical); data: synthetic across noise/channel; eval: accuracy, ROC AUC, confusion matrix; limits: theoretical validation only; future: RT optimization in multi-user QKD networks \\
    \bottomrule
  \end{tabular}
\end{table*}

\section{ML for QKD Key Metrics Prediction}

Besides being used for parameter optimization, ML has been applied to predict key performance metrics in QKD, including SKR, QBER, final key length (FKL) \footnote{In QKD, the \emph{final key length} indicates the number of secure bits remaining after error correction and privacy amplification on the raw key, reflecting the effective, usable secret key material that can be used for encryption or authentication while ensuring composable security under the given security parameter.}, and Strehl ratio (SR) \footnote{The fraction of optical power in the diffraction-limited core relative to a perfectly corrected system quantifies how much of the transmitted optical power is concentrated within the ideal diffraction-limited spot, serving as a metric of optical system performance under wavefront distortions and turbulence.}. Traditionally, calculating SKR, FKL, and QBER is based on computationally intensive convex optimization and iterative methods, which become burdensome when adapting to varying experimental conditions. Recent studies demonstrate that ML models, once trained on large datasets generated using these numerical methods, can efficiently generalize and infer SKR and QBER under dynamic experimental scenarios. However, directly predicting these key metrics using ML does not inherently guarantee security, requires redesigned loss functions and advanced data preprocessing to align with the unconditional security requirements of QKD systems, making this an active and promising area for practical QKD deployment. \cite{zhou2022neural}. 

In \cite{al2024towards}, the authors employed an autoencoder NN to predict the QBER and FKL with greater than 99\% accuracy. Predicting QBER enables the system to dynamically adjust error correction parameters in real-time, ensuring efficient, reliable, and secure key generation while adapting to varying noise conditions during quantum transmission between Alice and Bob, directly influencing the efficiency, reliability, and security of key generation. Similarly, predicting FKL allows the system to retain the maximum possible secure key length with minimal computational overhead under finite-size constraints. Recently, the authors in \cite{zhou2022neural} proposed ANNs to predict these parameters for homodyne detection discrete-modulated CV-QKD.  Their model achieved secure SKR predictions with 99.2\% reliability and achieved a speedup of 6–8 orders of magnitude over traditional numerical methods,  enabling real-time, efficient CV-QKD deployment. 
In \cite{liu2022automated}, the selected NN, using Bayesian optimization, reached 99.15\% in the quadrature phase-shift-keying (QPSK) heterodyne detection protocol and 99.59\% in the QPSK homodyne detection protocol. The model was further extended to achieve a speedup of about 107 times compared to traditional numerical approaches, thus enabling real-time SKR prediction. It also deployed an ML-enhanced Cascade protocol \cite{10.1007/3-540-48285-7_35} that uses two-way
 error-correction protocols for QKD and thus allows it to scale for high-speed applications while still providing security guarantees. Additionally, in \cite{ismail2019integrating}, the authors used ML techniques to enhance the characterization of free-space quantum channels by predicting the effects of distortions from atmospheric turbulence on quantum channels using the SR as a key metric.  By applying an RF Regressor, a Mean Absolute Percentage Error (MAPE) of 4.44\% was achieved, demonstrating high accuracy in turbulence prediction. Moreover, the introduction of fidelity as an additional feature improved the prediction performance, reducing the MAPE to 3.86\%. These findings suggest that ML can significantly enhance the real-time monitoring and optimization of free-space quantum communication, paving the way for more robust QKD systems and future applications in satellite-based quantum networks.

 In addition to direct prediction of QBER, SKR, FKL, and SR, a recent study \cite{ajlouni2025enhancing} applied ML models, RF and XGBoost, to predict the efficiency of QKD protocols under various operational conditions. The study evaluated three protocols: Adaptive, Entanglement-Based, and Hybrid, using simulated datasets with features such as key length, error rate, noise level, and channel stability. The ML models were trained post-simulation and used for efficiency prediction, not protocol execution. Although ML models did not influence protocol execution, they offered a powerful post-hoc analysis tool, enabling real-time inference of protocol performance under new or unseen conditions, enhancing decision-making for QKD deployment strategies.

\section{Conclusion}
\label{sec:conclusion}
QKD offers unconditionally secure communication rooted in quantum mechanics, yet its practical implementation remains challenged by noise, hardware imperfections, and environmental variability. This review examined how ML techniques are being used to enhance the security, adaptability, and performance of QKD systems under real-world constraints.

We structured our analysis around five primary categories: parameter optimization, network optimization, attack detection, protocol selection, and key metrics prediction. We reviewed how ML has been applied to fine-tune critical parameters across multiple layers of QKD systems. These included quantum signal calibration, polarization drift compensation, phase stabilization, post-processing refinement, and modulation variance adjustment. We highlighted models such as neural networks, random forests, LSTM architectures, and support vector machine, which improve key performance metrics, such as SKR and QBER. These approaches demonstrated substantial gains in system efficiency, with some eliminating the need for manual recalibration and others enabling near real-time operation.

Furthermore, we explored broader use cases of ML extending beyond physical-layer tuning. These include attack detection using supervised and unsupervised models, adaptive protocol selection based on channel or system conditions, key performance prediction for proactive system adjustment, and intelligent management of quantum communication networks. Our review showed how ML enables QKD systems to respond autonomously to dynamic threats, manage routing and resources, and improve security monitoring across both CV-QKD and DV-QKD platforms. These higher-level integrations indicate a trend toward intelligent and self-optimizing quantum networks.

We reviewed how a diverse range of ML techniques are used in QKD. Deep learning methods, such as DNNs, CNNs, LSTMs, and autoencoders, were frequently used for signal optimization and error correction. Classical ML models, including random forests, support vector machines, k-nearest neighbors, naïve Bayes, and XGBoost, were effective for protocol selection, performance prediction, and anomaly detection. Advanced methods such as reinforcement learning, Bayesian optimization, and generative adversarial networks supported adaptive decision-making and control strategies. Collectively, these approaches addressed challenges including noise reduction, side-channel defense, protocol switching, and resilience in dynamic environments.
The studies showed that ML integration led to improved accuracy in attack detection (up to 98 percent), reduced QBER, faster adaptation to environmental changes, and more reliable long-term system operation. These findings demonstrate that ML is not only a complementary tool but a foundational enabler for practical, secure, and scalable QKD deployments.

\subsection{Challenges}

Despite the promising outcomes discussed in this review, several challenges still affect the full integration of ML into practical QKD systems. A major challenge is the computational overhead associated with many ML models, particularly the deep learning ones, which can be resource-intensive and unsuitable for deployment in time-critical or low-power QKD environments. Additionally, most ML approaches are highly dependent on the availability and quality of training data, much of which is generated through simulations rather than real-world experiments. This raises concerns about model generalizability when exposed to actual deployment conditions, including noise, instability, and hardware variation.

Scalability also presents a critical constraint. While ML has shown strong performance in small-scale or single-link QKD setups, its application across large, distributed quantum networks with varying protocols and hardware configurations remains limited. Furthermore, security concerns surrounding the ML models themselves, such as susceptibility to adversarial attacks or model poisoning, are rarely addressed in the existing literature. Finally, the lack of standardization in evaluation metrics and model comparison makes it difficult to draw universal conclusions or establish reliable benchmarks between the different optimization approaches.

\subsection{Open Problems}

The community should focus on improving the practicality, robustness, and generalizability of ML-enhanced QKD systems. We motivate further research in the following open problems: 
\begin{itemize}
    \item Develop lightweight and efficient ML models that can operate in real time on constrained quantum hardware is essential for deployment in mobile, embedded, or high-speed QKD applications. This includes exploring model compression techniques, edge computing frameworks, and adaptive architectures that reduce inference time without compromising accuracy.
    \item More real-world validation is needed using experimental datasets that reflect diverse operational conditions, including fiber, satellite, underwater, and free-space QKD environments. Hybrid ML approaches that combine supervised, unsupervised, and reinforcement learning may offer better adaptability to varying network states and security threats. In particular, online learning methods could enable continual adaptation as environmental conditions evolve.
    \item Extending ML applications to underexplored areas such as integrated protocol control, trust management, and quantum network optimization would further increase the utility of ML in end-to-end QKD systems.   
    \item Collaboration between quantum communication and ML communities is vital for defining shared benchmarks, standardized datasets, and evaluation frameworks. 
    \item Exploring quantum machine learning (QML) algorithms to accelerate model training and inference, enable architectures like QCNNs, QLSTMs, and QSVMs, and directly exploit quantum features such as superposition and entanglement for key tasks in QKD, from parameter optimization to attack detection.

\end{itemize}

These steps are crucial to transition ML-enhanced QKD from isolated proofs-of-concept to fully deployable components of quantum-secure infrastructures.

\bibliography{sample}
\bibliographystyle{unsrt}

\end{document}